\documentclass[aps,prl,reprint,twocolumn,superscriptaddress]{revtex4-1}

\usepackage{graphicx}
\usepackage{epstopdf}
\usepackage{amssymb}
\usepackage{array}
\usepackage{amsmath}
\usepackage{bm}
\usepackage{color}
\usepackage{tabularx}
\usepackage[hidelinks]{hyperref}
\usepackage[normalem]{ulem}
\usepackage{euscript}
\usepackage{epsfig,psfrag,subfigure}
\usepackage{color}
\usepackage{amsfonts}
\usepackage{exscale}
\usepackage{amsbsy}
\usepackage{stmaryrd}
\usepackage{wasysym}
\usepackage{verbatim}
\usepackage{underscore}
\usepackage{xcite}
\usepackage{xr}
\usepackage[english]{babel}
\usepackage{url}
\usepackage{chngcntr}
\begin{document}


\title{Anomalous Thermal Diffusivity in Underdoped YBa$_2$Cu$_3$O$_{6+x}$}

\author{Jiecheng. Zhang}
\affiliation{Department of Physics, Stanford University, Stanford, CA 94305}
\affiliation{Geballe Laboratory for Advanced Materials, Stanford University, Stanford, CA 94305}
\author{E. M. Levenson-Falk}
\affiliation{Department of Physics, Stanford University, Stanford, CA 94305}
\affiliation{Geballe Laboratory for Advanced Materials, Stanford University, Stanford, CA 94305}
\author{B. J. Ramshaw}
\affiliation{Mail Stop E536, Los Alamos National Labs, Los Alamos, NM 87545}
\author{D. A. Bonn}
\affiliation{Department of Physics $\&$ Astronomy, University of British Columbia, Vancouver, British Columbia, Canada V6T 1Z1}
\affiliation{Canadian Institute for Advanced Research, Toronto, Ontario, Canada M5G 1Z8}
\author{R. Liang}
\affiliation{Department of Physics $\&$  Astronomy, University of British Columbia, Vancouver, British Columbia, Canada V6T 1Z1}
\affiliation{Canadian Institute for Advanced Research, Toronto, Ontario, Canada M5G 1Z8}
\author{W. N. Hardy}
\affiliation{Department of Physics $\&$  Astronomy, University of British Columbia, Vancouver, British Columbia, Canada V6T 1Z1}
\affiliation{Canadian Institute for Advanced Research, Toronto, Ontario, Canada M5G 1Z8}
\author{S. A. Hartnoll}
\affiliation{Department of Physics, Stanford University, Stanford, CA 94305}
\author{A. Kapitulnik}
\affiliation{Department of Physics, Stanford University, Stanford, CA 94305}
\affiliation{Geballe Laboratory for Advanced Materials, Stanford University, Stanford, CA 94305}
\affiliation{Department of Applied Physics, Stanford University, Stanford, CA 94305}
\affiliation{Stanford Institute for Materials and Energy Sciences, SLAC National Accelerator Laboratory, 2575 Sand Hill Road, Menlo Park, California 94025, USA}

\date{\today}






\begin{abstract}
The thermal diffusivity in the ab plane of underdoped YBCO
crystals is measured by means of a local optical technique in
the temperature range of $25 - 300$ K. The phase delay between a
point heat source and a set of detection points around it allows
for high-resolution measurement of the thermal diffusivity and
its in-plane anisotropy. Although the magnitude of the diffusivity
may suggest that it originates from phonons, its anisotropy
is comparable with reported values of the electrical resistivity
anisotropy. Furthermore, the anisotropy drops sharply below the
charge order transition, again similar to the electrical resistivity
anisotropy. Both of these observations suggest that the thermal
diffusivity has pronounced electronic as well as phononic character.
At the same time, the small electrical and thermal conductivities
at high temperatures imply that neither well-defined electron
nor phonon quasiparticles are present in this material. We
interpret our results through a strongly interacting incoherent
electron-phonon ``soup'' picture characterized by a diffusion constant
$D\sim v_B^2\tau$ where $v_B$ is the soup velocity, and scattering of
both electrons and phonons saturates a quantum thermal relaxation
time $\tau \sim \hbar /k_BT$.
\end{abstract}





\maketitle


The standard paradigm for transport in metals relies on the
existence of quasiparticles. Electronic quasiparticles conduct
electricity and heat. Phonon quasiparticles, the collective excitations
of the elastic solid (here, we discuss acoustic phonons)
also conduct heat. Transport coefficients, such as electrical and
thermal conductivities, can then be calculated using, for example,
Boltzmann equations\cite{Ziman1960}. However, such an approach fails when the quasiparticle mean free paths becomes comparable to the quasiparticle wavelength. For electrons it is the Fermi wavelength \cite{EmeryKivelson1995,Gunnarsson2003,Hussey2004}, while for phonons it is the larger of the interatomic distance or minimum excited-phonon wavelength \cite{Slack1979,Allen1994}. Understanding transport in non-quasiparticle regimes requires a new framework and has become a subject of intense theoretical effort in recent years triggering an urgent need for experimental results which can shed light on such regimes. In particular, in \cite{Hartnoll2015}, the diffusivity was singled out as a key observable for incoherent non-quasiparticle transport, possibly subject to fundamental quantum mechanical bounds.

In this letter we report high-resolution measurements of the thermal diffusivity of single crystal underdoped  YBCO$_{6.60}$ (an ortho-II YBa$_2$Cu$_3$O$_{6.60}$),  and YBCO$_{6.75}$ (an ortho-III YBa$_2$Cu$_3$O$_{6.75}$). We are particularly interested in the anisotropy of the thermal diffusivity as measured along the principal axes $a$ and $b$ (which is the chain direction) in the temperature range 25 to 300 K. We use a non-contact optical microscope to perform local thermal transport measurements on the scale of $\sim$10 $\mu$m, hence avoiding inhomogeneities, particularly twinning and grain boundary effects. Our principal experimental results are: i) the measured thermal diffusivity is consistent with the previously measured thermal conductivity and specific heat, satisfying $\kappa = c D$. The high-temperature specific heat is known to be dominated by phonons. However, ii) the phonon mean free path implied by the magnitude of the measured diffusivity is of order, or smaller than, the phonon wavelength. In addition:  iii) the intrinsic thermal anisotropy is found to be almost identical to the electrical resistivity anisotropy; and iv) the thermal anisotropy starts to decrease rather sharply below the charge order transition. 

A complete understanding of transport in the high temperature regime of the YBCO (or similar) material system requires that we interpret our diffusivity results  together with  previously reported measurements of the charge sector on similar crystals at temperatures above the charge order transition, primarily photoemission spectroscopy \cite{Norman1997,Lanzara2001} and electrical resistivity  \cite{Segawa2001,Ando2002,LeBoeufThesis}. Those measurements suggest that the electronic mean free path is also comparable or smaller than the electron wavelength, and thus at or beyond the Mott-Ioffe-Regel (MIR) limit  [The MIR limit has been expressed in different ways in the literature, e.g. as $k_F\ell \approx 1$ or $\ell/a\approx 1$ ($k_F$ is the Fermi wavevector, $a$ is the lattice constant, and  $\ell$ is the mean free path). These approaches typically produce the same order of magnitude estimate. In this paper we use the criterion proposed in \cite{EmeryKivelson1995} of $\ell/\lambda_F \approx 1$ where $\lambda_F = 2\pi/k_F$].  The simultaneous destruction of phonon and electron quasiparticles, together with comparable anisotropies in thermal and charge transport, suggest that scattering is dominated by  strong electron-phonon interaction. The lack of coherent response is furthermore incompatible with electron-phonon composite quasiparticles such as polarons \cite{Emin1992} or bipolarons \cite{Alexandrov1994}. We are therefore led to conjecture  a  novel type of transport in the YBCO (or similar) system, in the temperature regime where quasiparticles are not well defined, which is dominated by diffusion of an electron-phonon ``soup''. Furthermore, all relaxation processes of electrons and phonons, and hence of charge and heat, are saturated at the thermal relaxation time $\tau \sim \hbar/k_BT$. This `Plankian' timescale has previously been proposed to underpin transport across many families of unconventional metals\cite{subirbook,Zaanen2004,ScatterSimilar}, and is widely observed in optical data on cuprates \cite{PhysRevB.42.6342,PhysRevB.41.11237, Liu99, Marel2003}. A simple model based on the above ansatz shows excellent fit to the temperature dependence of the measured thermal diffusivity using measured material parameters. This approach generalizes the recently proposed incoherent metallic transport  \cite{Hartnoll2015} to the case where the electronic system exhibits strong interactions with additional degrees of freedom\textemdash the phonons. An immediate further generalization of our idea could apply to complex insulators at high fields, with magnetic or polarization excitations playing the role of the phonons.

\section*{Results and Discussion}
\subsection{Thermal Diffusivity Measurements}
For the high resolution thermal diffusivity measurements we use a  photothermal microscope (see See SI - Supplemental Information), where a modulated power of a heating laser beam causes ripples in the temperature profile at the sample surface, which may be measured by a probing laser due to the change in reflectivity as a function of temperature. We obtain the diffusivity $D$ by fitting the phase delay between the source and the response signals as a function of the modulation frequency $\omega$ at fixed distance $r$ between the source and probe beam  (see Methods).  In the case of an anisotropic sample, the extracted diffusivity depends on orientation $\theta$ as follows
\begin{equation}
\label{anisoD}
	D=\frac{D_aD_b}{D_b\cos^2\theta+D_a\sin^2\theta}
\end{equation}
where $D_1$ and $D_2$ are the diffusivities of the two principle axes on the surface. Since $D=\kappa/c$, where the specific heat $c$ is a scalar quantity, the diffusivity inherits its spatial anisotropy from the thermal conductivity tensor $\kappa$.
\begin{figure}[h]
	\centering
	\includegraphics[width=\linewidth]{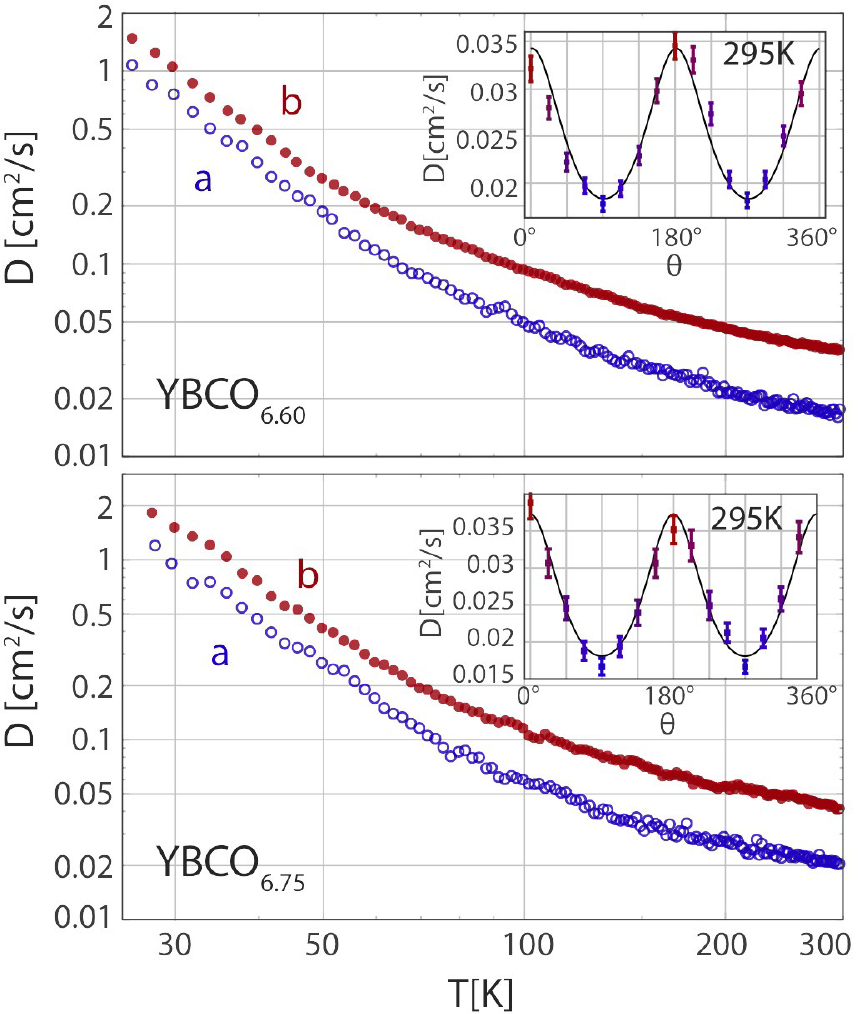}
	\caption{\label{DvsT} (color online) Thermal diffusivity of YBCO$_{6.60}$ and  YBCO$_{6.75}$ single crystals, extracted from phase measurement, plotted on a log-log scale as a function of temperature in the range $25-300$ K. Insets show diffusivity measured at room temperature as a function of orientation around the heating spot, with the solid lines representing fits to Eq.~\ref{anisoD} (see text).   Error bars are almost entirely due to uncertainty in determining lasers spots separation.}
\end{figure}
Several YBa$_2$Cu$_3$O$_{6+x}$  samples were measured; all showed consistent results with the two samples reported here: a detwinned single crystal of YBCO$_{6.60}$ measuring approximately 3 mm $\times$ 2 mm $\times$ 1 mm, and a detwinned single crystal of YBCO$_{6.75}$ measuring approximately 2 mm $\times$ 1 mm $\times$ 0.4 mm. Fig.~S1(c) in SI shows the surface of the YBCO$_{6.75}$ with the focused laser spots under polarized light, where bright/dark stripes are the twin domains. The small scale of measurement enables us to measure local diffusivity in all directions of the Cu-O planes while avoiding edges, twin boundaries, and other visible defects when possible. At room temperature (RT), both samples have reflectivity $R\approx 0.15$, and $dR/dT\approx10^{-4}$ K$^{-1}$ at 820 nm. The typical power of the probing laser is $\lesssim 0.2$ mW, and typical RMS power of the heating laser is $\lesssim 0.5$ mW. Using thermal conductivities reported in existing literature \cite{YBCOConductivity}, we estimate the increase in mean temperature due to both lasers to be less than 1 K in the entire temperature range of interest. To check the alignment of the sample, we first measure diffusivity as a function of sample orientation relative to the axis of displacement of the laser spots, shown in the insets of Fig.~\ref{DvsT}.  Error bars are due to the $\sim 0.5$ $\mu$m uncertainty in determining the distance $r$ between the two spots and thus in extracting $D$ from the data.  The solid line is a fit to Eq.~\ref{anisoD}, showing excellent agreement.  The offset angle is left as a free parameter, but agrees with the alignment to the sample edge to within $1^\circ$.  We find that the local orientation of the $a$- and $b$-axis swap between different twinning domains, verifying the single-domain nature of or measurements. Small variations (of order 10\%) in diffusivity are measured at different areas on the sample surface, which we attribute to material inhomogeneity. Once the principal axes are determined from the detailed anisotropy study, the temperature dependence of the diffusivity is measured along each of the principal axes in a continuous temperature sweep at a fixed frequency. 

\subsection{Initial Observations}
\begin{figure}[h]
	\centering
	\includegraphics[width=\linewidth]{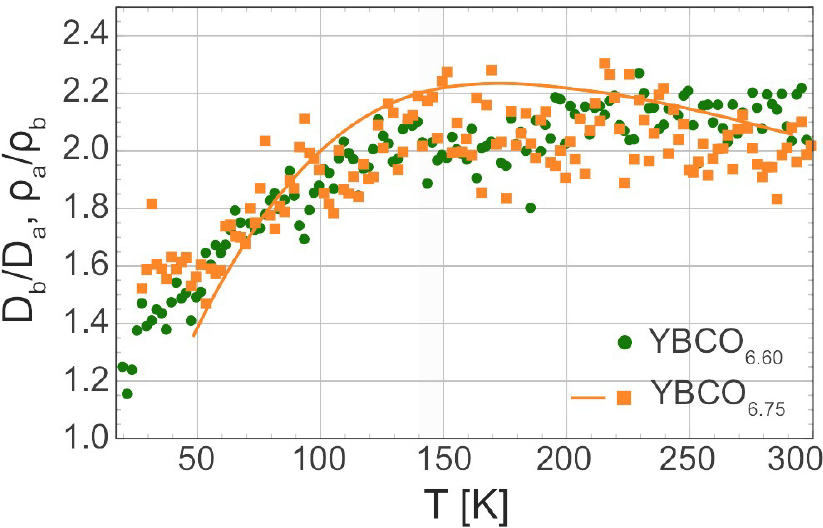}
	\caption{\label{anisoT} (color online) Anisotropy of the $ab$-plane thermal diffusivity as a function of temperature of YBCO$_{6.60}$ (green circles) and YBCO$_{6.75}$ (orange squares). Charge density order occurs at around $140-150$ K in both materials (see e.g. \cite{Cyr2015}), marked by the grey region. Note that anisotropies decrease significantly below the transition, signifying the non-trivial role the electronic system plays in the thermal transport. Solid line is the electrical anisotropy in the $ab$ plane on similar crystals adopted from ~\cite{LeBoeufThesis}.}
\end{figure}

The temperature dependence of the diffusivities along both the $a$- and $b$-axis for the two samples are shown in Fig.~\ref{DvsT}. We observe that the diffusivity increases at lower temperatures, increasing by almost two orders of magnitude in the temperature range studied here.  Using existing measurements of specific heat on similar YBCO crystals \cite{YBCOcapacity}, we obtain the respective thermal conductivities of the crystals; both show excellent agreement with previously measured thermal conductivities of YBCO crystals with similar doping (e.g. $\kappa_a$ for YBCO$_{6.60}$ was previously measured by Waske {\it et al.} and Minami {\it et al.} \cite{YBCOConductivity,Minami2003}). The temperature dependence of the thermal anisotropy, $D_a/D_b$, for both samples are shown in Fig.~\ref{anisoT}. Three main features of the data are observed. First, the anisotropies of the two samples are similar. Second, the anisotropy is almost temperature independent with $D_a/D_b \sim 2$ at high temperatures, but decreases sharply below the charge density wave (CDW) transition ($\sim140-150$ K \cite{Cyr2015}). Third, the size and temperature dependence of the thermal anisotropy are very similar to those of the anisotropy of the electrical resistivity. 
In Fig.~\ref{anisoT} we also plot the resistivity anisotropy of YBCO$_{6.75}$ from ref.~\cite{LeBoeufThesis}, measured on very similar crystals.  Since the resistivity measures transport of the electronic system, we conclude here that the thermal diffusivity exhibits a strong electronic character. Furthermore, while at low doping levels the charge order is strongly anisotropic (see e.g. \cite{Blanco2014}),  for the YBCO$_{6.60}$ \cite{Ghiringhelli2012} and YBCO$_{6.75}$ \cite{Achkar2012} reported here, an almost isotropic CDW transition would not strongly affect the diffusivity anisotropy of conventional phonons. Instead, in mirroring the behavior of the electrical resistivity, the decrease in anisotropy  again indicates electronic contribution to the thermal diffusivity. The onset of CDW order can both change the scattering mechanism of the electrons and also lead to new collective transport dynamics. The electronic excitations may ultimately regain their quasiparticle character, including possible polaronic behavior \cite{Emin1992,Alexandrov1994}. Furthermore, the CDW transition in this material is correlated with a strong electron-phonon interaction \cite{bonnoit2012probing, LeTacon2014}. We further elaborate on these points below.

\subsection{Good Metals vs.~Bad Metals}
The conventional treatment of heat conduction in itinerant solids assumes the existence of well defined quasiparticle excitations that transport entropy: electrons and phonons.  The mean free paths, $\ell_e$ and $\ell_{ph}$ for the electrons and phonons respectively, are assumed to be much larger than their respective wavelength. For electrons to be well defined quasiparticles we require that $\ell_e /\lambda_F\gg 1$,  where $\lambda_F$ is the Fermi wavelength. The limiting case $\ell_e/\lambda_F\sim 1$ is called the Mott-Ioffe-Regel (MIR) limit \cite{MIR}, beyond which the material is dubbed a ``bad metal'' \cite{EmeryKivelson1995}.  Analogously, for well defined phonon excitations we require that $\ell_{ph} \gg {\rm max}\{a,\lambda_{min}\}$, where $a$ is the lattice constant and   $v_s/\lambda_{min}(T)$ ($v_s$ is the sound velocity) is the highest acoustic phonon frequency excited at temperature $T$ \cite{Slack1979,Allen1994}. If both conditions are met, the thermal conductivity of the solid will be the sum of the thermal conductivities of the electrons and phonons, which can also be written as products of the respective diffusivities and specific heat capacities:
\begin{equation}
\label{dd}
\kappa = \kappa_e + \kappa_{ph}=c_eD_e + c_{ph}D_{ph} = cD \,.
\end{equation}
The last equality states that the measured thermal diffusivity is a heat capacity weighted average of the two diffusivities, and $c=c_e+c_{ph}$. 

\begin{table*}[ht]
\begin{tabular}{cccccccccccc}
\hline
Metals& D & c &$\kappa^*=cD$ &$\kappa$ & $\rho$ & $v_s$ & $\kappa_{ph}^{min}\approx cv_sa$ & $\kappa_e(L_0)$ &$ L/L_0$ &$\theta_D$ \\
&cm$^2$/s & J/cm$^3$-K & W/cm-K &W/cm-K & $\times10^5\mu\Omega$-cm  & cm/s & W/cm-K & W/cm-K & &K \\
  \hline
Copper    &1.15$^4$    & 3.45$^1$   &	3.97 & 3.98$^1$  &1.67$^1$  & 3.56$^1$   &0.044  &4.4   & 0.91& 310$^2$ \\ 
Gold    & 1.28$^5$   & 2.5$^1$   	&3.2 &3.15$^1$  &2.24$^1$ & 3.25$^1$   & 0.033    & 3.3 & 0.96& 185$^2$ \\ 
Silver    & 1.61$^4$    & 2.47$^1$  	&4.0 &4.27$^1$ &  1.59$^1$& 3.6$^1$   & 0.036  & 4.6 & 0.95 & 220$^2$\\ 
Tungsten   &  0.68$^6$   & 2.56$^1$  	&1.74 &1.78$^1$  & 5.65$^1$& 5.25$^1$   & 0.042  & 1.3 & 1.27& 315$^2$ \\ 
  &  &    &  	&  &  &    &    & &  & \\ 
Hg   &   0.043$^7$  & 1.88$^1$  &	0.081& 0.084$^1$  & 96$^2$&1.45$^1$ & 0.0085  & 0.074 & 1.11 & 110$^2$\\  
Constantan    &  0.064$^{8}$   & 3.65$^2$&0.23	& 0.21$^{1,2}$ & 49.9$^1$& 5.2$^1$  &  0.07  & 0.15 & 1.32 &$390^{3}$ \\ 
Inconel 718    &  0.029$^{23}$   & 3.5$^2$& 0.101   	& 0.097$^2$   & 156$^2$&4.94$^2$ &  0.06  & 0.05 & 1.56 & $410^{3}$ \\  
       &    &  	&    &    &  &  & &  && \\ 
YBCO$_{6.6}$($a$-dir)   & 0.016-0.018$^{24}$    & 2.7$^{9}$  &0.043-0.05	& 0.05-0.065$^{12}$   &560-700$^{13}$ &6.05$^{18}$   & 0.063   & 0.013 & 5.0-5.4 & 410$^{11}$\\ 
 YBCO$_{6.75}$($a$-dir)   & 0.018-0.02$^{24}$    & 2.7$^{9}$ & 0.05-0.054	& 0.047-0.068$^{12}$   &430-500$^{13}$ & 6.05$^{18}$   & 0.063   & 0.017 & 4.1-5.0& 410$^{11}$ \\ 
LSCO(x=0.13)   &  0.021$^{21}$  & 2.67$^{10}$  &0.056	& 0.057$^{14}$  &700$^{16}$& 5.9$^{19}$   & 0.06   & 0.01 & 4.5-5.5 & 400$^{10}$\\ 
BSCCO:2212   &  0.021$^{22}$   & 2.35$^{11}$ &0.048	& 0.058$^{15}$   &580-780$^{17}$ &4.3$^{20}$  & 0.055   & 0.009 & 4.6-6.2& 280$^{11}$\\ 
\hline
\end{tabular}
 \caption{{\bf Room temperature thermal transport parameters for selected good metals, disordered metals, and cuprates.}  \newline Comparison of room-temperature thermal properties of good metals, disordered metals, and cuprates. LSCO data are for La$_{1.87}$Sr$_{0.13}$CuO$_4$ (see Fig. S4). BSCCO data
are for optimally doped Bi$_2$Sr$_2$CaCu$_2$O$_8$ as found in the literature. YBCO$_{6.60}$(a-dir) and YBCO$_{6.75}$(a-dir) a-direction data are on single crystals from similar doping, whereas diffusivity data are from this work. Discussions on $\kappa^*=cD$, $\kappa_{ph}^{min}\approx cv_s\ell_{ph}$ (with $\ell_{ph}=a$; $a$ is the lattice constant), and $\kappa(L_0)=L_0T/\rho$= are in the text. Superscripts are the references for the quoted data as follows:  $^1$=\cite{CRC1985}; $^2$=\cite{Ekin2006}; $^{3}$=calculated from~\cite{Ekin2006}; $^4$=\cite{Parker1961}; $^5$=\cite{Yunus2001}; $^6$=\cite{Hofmann2015}; $^7$=\cite{Ang1974}; $^8$=\cite{Sundqvist1992}; $^9$=\cite{Junod1987,YBCOcapacity}; $^{10}$=\cite{Junod1987}; $^{11}$=\cite{Junod1994}; $^{12}$=\cite{Inyushkin1996,Minami2003};  $^{13}$=\cite{Segawa2001,Minami2003,LeBoeufThesis}; $^{14}$=\cite{Nakamura1991}; $^{15}$=\cite{Allen1994}; $^{16}$=\cite{Ando1995}; $^{17}$=\cite{Fastampa2003,Triscone1996};$^{18}$= \cite{Lei1993}; $^{19}$=\cite{Sarrao1994}; $^{20}$=\cite{Saunders1994}; $^{21}$=see SI; $^{22}$=\cite{Wu1993}; $^{23}$=\cite{Sweet1987}; $^{24}$=this work (spread due to inhomogeneity).}
 \label{tt}
\end{table*}
The consequences of Eqn.~\ref{dd} are well demonstrated in the case of  both good  and disordered metals, even at temperatures of order the Debye temperature. Table~\ref{tt} gives  examples of room temperature (RT) thermal transport parameters for good metals (copper, gold, silver, tungsten), metals with large RT resistivity (mercury, constantan, Inconel), and several cuprates, which are known to be ``bad metals'' at RT. In the case of good and disordered metals, ample data is available on identically prepared samples to confirm that at RT the thermal conductivity $\kappa$ closely matches the measured diffusivity times the measured specific heat ($\kappa^*$ in Table~\ref{tt}). Furthermore, when the electronic thermal conductivity $\kappa_e(L_0)$ is calculated from the resistivity using the Wiedemann-Franz law ($\kappa \rho/T=L$) and the theoretical value of the Lorenz number $L_0=2.44\times10^{-8}$W$\cdot\Omega/$K$^2$, the result is very close to the measured thermal conductivity, yielding a measured $L/L_0 \approx 1$.

Unfortunately, not much data is available for the cuprates where the diffusivity, specific heat, thermal conductivity, and resistivity have all been measured on the same sample, or at least on same-doping samples made with the same protocol. For example, $a$-direction thermal conductivity on YBCO$_{6+x}$ was reported by Minami {\it et al.} \cite{Minami2003}, but for similar dopings, their crystals show  much larger resistivity than measurements done on crystals similar to ours \cite{LeBoeufThesis}.  At the same time, Inyushkin {\it et al.} measured twinned crystals with similar doping levels, which are expected to yield a larger value as the $a$-and $b$- directions average \cite{Gold1994}. We therefore base our estimates  on a range of thermal-parameters values as found in the literature. 

Analyzing the available data for the cuprates, we first note that also here $\kappa\sim\kappa^*$, but now $\kappa \gg \kappa_e(L_0)$. At the same time we find it to be very close to, and sometimes smaller than, $\kappa_{ph}^{min}$. This minimum phonon thermal conductivity is calculated as the product of the sound velocity, the lattice constant and the specific heat, and amounts to setting $\ell_{ph} = a$ as discussed above. It is a lower bound to the total thermal conductivity of a system with well defined quasiparticles regardless of electron participation (a complementary bound on $\kappa_{ph}^{min}$ using $\lambda_{min}$ instead of $a$ is discussed below.) For good metals $\kappa_{ph}^{min} \ll \kappa$,  and  $L/L_0\sim1$  as is expected from electron-dominated thermal transport. Mercury at RT is a liquid with a relatively slow sound velocity. Nevertheless, it shows similar behavior to the best metallic elements with thermal transport dominated by electrons. On the other hand, with increasing resistivity due to disorder, constantan and Inconel show a tendency to a decreased thermal conductivity, and an increasing Lorentz number. The bad metal cuprates on the other hand show at RT $\kappa \approx \kappa_{ph}^{min}$, with anomalously large Lorentz number. These facts  might raise doubts on whether high-temperature thermal transport in cuprates has any significant electronic contribution at all, cf. \cite{Minami2003}, in apparent tension with the electronic character noted in the previous section.

\subsection{Failure of the Quasiparticle Picture}
While at lower temperatures it is believed that quasiparticle excitations are well defined \cite{Norman1997,Deng2013}, the situation must change close to the MIR limit. Indeed, applying the conventional treatment to our diffusivity measurements in the temperature range $\gtrsim$ 150 K raises several problems that challenge the self-consistency of the quasiparticle approach. Within a quasiparticle interpretation, $\kappa_e(L_0) \ll \kappa$, and so we would conclude that the high-temperature thermal transport is dominated by phonons. Even without assuming quasiparticles, at high tempeature the phonon specific heat is overwhelmingly large compared to that of the electrons \cite{YBCOcapacity,strongspecificheat}. Thus, since the product of the  specific heat and our measured diffusivity yields the commonly measured thermal conductivity, the naive conclusion is that phonons dominate the high temperature thermal transport, and we should expect to measure $\kappa = cD \approx c_{ph}D_{ph} =\kappa_{ph}$.  However, even in this phonon-dominated picture, we see that $\kappa_{ph} \approx \kappa_{ph}^{min}$, and so $\ell_{ph} \approx a$.  Since room temperature is well below the Debye temperature for this material, it is appropriate to use a stricter bound on $\kappa_{ph}^{min}$ associated with $\lambda_{min}$ (instead of $a$).  Assuming that the highest frequency phonon excited at RT is proportional to temperature, the corresponding minimum phonon wavelength is $\lambda_{min}\sim h v_s/k_BT$. The sound velocities of a similar YBCO$_{6.60}$ crystal have been measured yielding  $v^a_s=6.0\times10^3m/s$ and $v^b_s=6.5\times10^3m/s$ \cite{Lei1993}. Shear sound velocity may be a factor $\sim\sqrt{3}$ smaller, which does not change the following arguments.  Note that these velocities  (along with the lattice constants) would yield a much more isotropic diffusivity  than the measured $D$.  Assuming now two-dimensional phonon transport, the ``classical'' phonon diffusivity satisfies $D_{ph}=v_s\ell_{ph}/2$, and thus we estimate $\ell_{ph}/\lambda_{min} \approx 2D k_BT/h v_s^2 \sim 0.6 < 1$ (for $D_a$ at RT), violating the condition for well defined phonon excitations. A similar analysis was previously reported by Allen {\it et al.} \cite{Allen1994} on Bi$_2$Sr$_2$CaCu$_2$O$_8$ single crystals, concluding that phonons are poorly defined in this system. However, their treatment of the electronic contribution relied on a standard application of the WF  law, which was subsequently shown by Zhang {\it et al.} \cite{Zhang2000} to fail around RT. Finally we note that (see table~\ref{tt}) the values of $a$-direction $D$ for the YBCO$_{6+x}$ crystals is practically the same as the values for the more isotropic (in the Cu-O planes)  cuprates. Similar observations can be found for the resistivity of the cuprates (e.g. $b$-axes resistivities can be found in~\cite{LeBoeufThesis,Segawa2001}). This must indicate that the chains have excess electronic conduction rather than phononic one.

The phonon quasiparticle approach is therefore not consistent. Likewise for the electrons, examination of the resistivity data for similar crystals at around RT yields $\ell_e/\lambda_F \sim 1$ (see e.g. \cite{Segawa2001,LeBoeufThesis}). Beyond the short mean free paths, the phonon dominance suggested by a quasiparticle interpretation is incompatible with the fact that our measured diffusivity anisotropy is similar to the electrical resistivity anisotropy, and sensitive to the onset of charge order at $\sim$150 K. We are led to conclude that \emph{thermal transport has strong electronic in addition to phononic character, with no simple way to separate them}, especially in view of the typically very strong electron-phonon interaction in the cuprates  \cite{Lanzara2001}. 

\subsection{The Case for an Incoherent Electron-Phonon ``Soup''}
We are therefore led to propose a new approach to transport in strongly interacting systems where neither elementary excitations are well defined. Without quasiparticles, including the absence of emergent well-defined electronic excitations (e.g collective modes related to a possible symmetry breaking, or dressed coherent excitations such as polarons and bipolarons), the mean free path has no meaning.  However, microscopic relaxation timescales can still be defined. Following \cite{Hartnoll2015}, we conjecture that all microscopic degrees of freedom, electronic or phononic, saturate a (momentum non-conserving) relaxation bound leading to overdamped diffusive transport with quantum thermal timescales $\tau\propto\hbar/k_BT$. The resulting diffusion coefficient is connected to the thermal timescale through a (temperature dependent) effective velocity, $v_B(T)$, such that
\begin{equation}
D =  \frac{1}{2} v_B(T)^2 \frac{\hbar }{k_B T}
\label{butterfly}
\end{equation}
where the factor of 2 represents the quasi two-dimensional diffusion in the Cu-O planes.  This approach suggests that in the strongly coupled, high temperature limit the electron-phonon system behaves as a composite, strongly correlated ``soup'' with an effective velocity $v_B$. This velocity is expected to lie between the faster Fermi velocity of the electrons and the much slower sound velocity of the phonons: $v_s < v_B < v_F$. 

To obtain an estimate of $v_B$, we may attempt to extrapolate the expression for the thermal conductivity from the regime where quasiparticles are well defined and equation~(\ref{dd}) holds
\begin{equation}
\kappa = cD= c_e\left(\frac{1}{2}v_F^2\tau_e\right) + c_{ph}\left(\frac{1}{2}v_s^2\tau_{ph}\right) 
\label{ddd}
\end{equation}
to the new strongly coupled regime. Assuming a smooth interpolation between the two regimes, we bound the electron and phonon relaxation times in the above equation by
\begin{equation}\label{tau}
\tau_{ph}=\alpha_{ph}\frac{\hbar}{k_BT}, \ \ \ \ \ \ and \ \ \ \ \  \tau_{e}=\alpha_{e}\frac{\hbar}{k_BT}
\end{equation}
where $\alpha_{ph}$ and $\alpha_{e}$ are numerical constants of order unity. The resulting expression for the diffusivity is
\begin{equation}
D = \frac{\hbar}{2k_BT} \left(\alpha_e \frac{c_e}{c}v_F^2 + \alpha_{ph} \frac{c_{ph}}{c}v_s^2\right) \ .
\label{composite}
\end{equation}
Equation~(\ref{composite}) now has the form~(\ref{butterfly}) with $v_B = \bar{v}_B(T)$, where
\begin{equation}
\bar{v}_B(T)^2 = \alpha_e \frac{c_e}{c}v_F^2 + \alpha_{ph} \frac{c_{ph}}{c}v_s^2 \ ,
\label{vb}
\end{equation}
which clearly satisfies the condition $v_s < \bar{v}_B < v_F$ (we note that,
in the expression for the thermal conductivity, only the heat
capacity associated with the modes that propagate in the measured
direction and carry entropy should be included. However,
this difference may only change the expression by a factor of
order unity). More generally one would like
to identify $v_B$ with a universally defined non-quasiparticle velocity. It has been noted \cite{Blake:2016wvh} that in some non-quasiparticle systems it is the ``butterfly velocity''  \cite{Shenker:2013pqa,Roberts:2014isa}  that appears in the diffusivity formula~\ref{butterfly}.

\begin{figure}
	\centering
	\includegraphics[width=\linewidth]{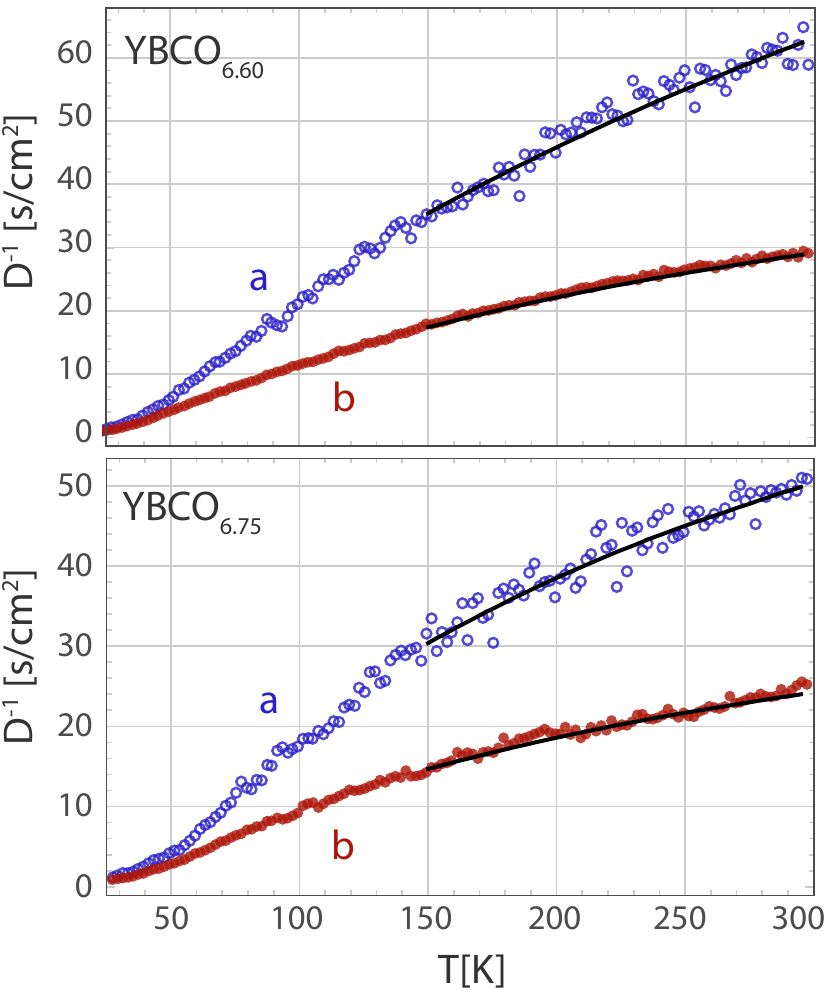}
	\caption{\label{fit} (color online) Inverse diffusivity along the $a$- (unfilled blue circles) and $b$-axis (filled red circles) for both materials.  The solid lines are fits to Eq.~\ref{composite} (see text.)}
\end{figure}

Fig.~\ref{fit} shows the  temperature dependence of $D^{-1}$ with the fit to Eqn.~(\ref{composite})  above 150K,  where we are free of the interference with the onset of charge order \cite{Ghiringhelli2012,Achkar2012}. Using the known temperature dependent total heat capacity  \cite{YBCOcapacity} and  electronic specific heat \cite{Moler1997}  (at RT $c_{ph}\sim c$ while $c_e\sim 0.014c$), and assuming in-plane Fermi velocity found e.g. in ARPES measurements $v_F\approx 2.15\times 10^{7}$cm/sec \cite{Fournier2010}, we find e.g. for YBCO$_{6.75}$ that in the Cu-O planes the prefactor $\alpha_e \approx 0.25 \pm 0.02$, and  $ \alpha_{ph} \approx 1.64 \pm 0.09$, while along the chains $\alpha_e\approx 0.53 \pm 0.02$, and  $\alpha_{ph} \approx 2.8 \pm 0.2$. Hence the temperature dependent microscopic velocity $\bar{v}_B$, defined in Eqn.~(\ref{vb}), exhibits significant character of both electrons and phonons  with similar coefficients of order unity and velocity $v_s < \bar{v}_B <  v_F$. The small errors associated with the coefficients (see Supporting Information (SI)) signal the robustness of the fit, while the independently measured temperature dependences of $c_e$ and $c_{ph}$ are responsible for the curvature away from $D^{-1} \propto T$ in the figure.
Furtheremore, signatures of an electron-phonon soup are seen in the behavior of the electrical resistivity, which is not exactly $T$-linear over our temperature range. Comparing our diffusivity data with existing electrical resistivity measurements \cite{Segawa2001,LeBoeufThesis} shows that in the bad metal regime $d\rho/dT \propto d (D^{-1})/dT$, implying that the sound speed contributes to the electrical resistivity.

The above analysis yields a novel transport mechanism for strongly interacting systems that exceed the quasiparticle mean free path limit. Thermal transport proceeds by collective diffusion of a composite electron-phonon ``soup'', which is distinct from any system exhibiting well defined electron-phonon quasiparticles such as polarons or bipolarons \cite{Emin1992,Alexandrov1994}. Entropy diffusion is characterized by thermal timescales and a unique velocity. Consequently, commonly used electronic-transport characteristics, such as e.g. the Wiedemann-Franz law, are not well-defined in this regime. In the YBCO cuprates this scenario seems to persist down to the charge order temperatures, below which electrons presumably start to regain their quasiparticle character and hence decouple from the electron-phonon ``soup.'' Obviously, a main ingredient of this scenario is the strong electron-phonon interaction which in turn may have some impact on the high $T_c$ found in the cuprates.

\subsection*{Conclusions}
In conclusion, we have shown that the underdoped YBa$_2$Cu$_3$O$_{6+x}$ system, above the charge order transition, exhibits anomalous thermal transport. Neither the phonons nor the electrons are well defined quasiparticles, while their strong mutual interactions cause both to saturate the relaxation timescale at $\sim\hbar/k_BT$. This results in a unique type of heat transport carried by an incoherent composite fluid, which we dub the an electron-phonon ``soup'', characterized by an effective velocity $v_s < v_B < v_F$.  We suggest that such behavior is ubiquitous in strongly interacting complex systems at high temperatures, and thus propose that it may explain much of the anomalous transport in ``bad metallic'' systems [e.g. catalogued by Bruin {\it et al.} \cite{ScatterSimilar}, and discussed in terms of spectral weight transfer in \cite{Gunnarsson2003,Hussey2004,Jaramillo2014}. Additional diffusivity measurements on these systems may test this proposal.

\section{Materials $\&$ Methods}
\subsection*{Samples} 
 YBa$_2$Cu$_3$O$_{6+y}$ single crystals were grown in non-reactive BaZrO$_3$ crucibles using a self-flux technique \cite{Liang1998}, with the Cu-O chains oxygen content accurately determined as described in Ref: \cite{Liang2006}. The crystals used in our experiments were de-twinned. No twinning domains can be observed on the surface of the YBCO$_{6.60}$ sample, and sparse thin strips of remnant domains can be seen on the YBCO$_{6.75}$, which were experimentally measured to have no effect on the measured diffusivity.

\subsection*{Diffusivity Measurements} 
For the high resolution thermal diffusivity measurements we use a home-built photothermal microscope \cite{FantonThesis,XDWu1993}, described in details in the SI.    The output power of a heating laser is modulated sinusoidally at a frequency $\omega\sim$ 1 to 50 kHz, much smaller than typical electronic equilibration time \cite{Li2015}, while a second laser measures the differential reflectivity at a fixed distance ($r$) from it. The diffusivity $D$ is obtained by fitting the thermal phase delay $\phi$ between the source and the response signals as a function of frequency $\omega$: $D = \omega r^2 / 2 \phi^2$.  Frequency sweeps at different distances yield consistent diffusivity values verifying the heat propagation model we used (for details see the SI).

\subsection*{Fits to Data}
We use Eqn.~\ref{composite} to fit the thermal diffusivity data. Literature values are used for the total specific heat \cite{YBCOcapacity}, the electronic specific heat  \cite{Moler1997}, Fermi velocity \cite{Fournier2010}, and sound velocities \cite{Lei1993,soundvelocity}. 


\section*{Acknowledgments}
We  thank Subir Sachdev, Sam Lederer and Steve Kivelson for many insightful discussions. Work supported by the Gordon and Betty Moore Foundation through the EPiQS Initiative, Grant GBMF4529, and by a Department of Energy Early Career Award (SAH). 

\section*{Significance Statement}
Transport in the so-called ``bad-metallic'' regime of strongly correlated electron systems, with no well-defined electronic quasiparticles, has been a long-standing challenge in theoretical physics.  This challenge has motivated the collection of an ample amount of data on bad metals. However, so far emphasis has been given to the charge sector, with the host crystal lattice treated as a well-defined phonon background. In this paper we show that for the cuprates, in the bad-metallic regime where the resistivity exceeds the ``Mott-Ioffe-Regel'' limit, phonon excitations are also not well-defined.  The data leads to a thermal transport scenario where entropy is carried by an overdamped diffusive fluid of electrons and phonons characterized by a unique velocity and a quantum-limited relaxation time $\hbar/k_BT$.

\pagebreak

\section*{SUPPORTING INFORMATION (SI)}

\setcounter{figure}{0}
\renewcommand{\figurename}{S}

\subsection*{Samples} 
 YBa$_2$Cu$_3$O$_{6+y}$ single crystals were grown in non-reactive BaZrO3 crucibles using a self-flux technique \cite{Liang1998}, with the Cu-O chains oxygen content accurately determined as described in Ref: \cite{Liang2006}. The CuO-chain oxygen content was set to $y = $ 6.60, 6.67, and 6.75 by annealing in a flowing O2:N2 mixture and homogenized by further annealing in a sealed quartz ampoule, together with ceramic at the same oxygen content. The absolute oxygen content ($y$) is accurate to $\pm$0.01 based on iodometric titration. The crystals used in our experiments were de-twinned. No twinning domains can be observed on the surface of the YBCO$_{6.60}$ sample, and sparse thin strips of remnant domains can be seen on the YBCO$_{6.75}$, which were experimentally measured to have no effect on the measured diffusivity.

\subsection{Principles of the Photothermal Apparatus}
\begin{figure}[ht]
	\centering
	\includegraphics[width=\linewidth]{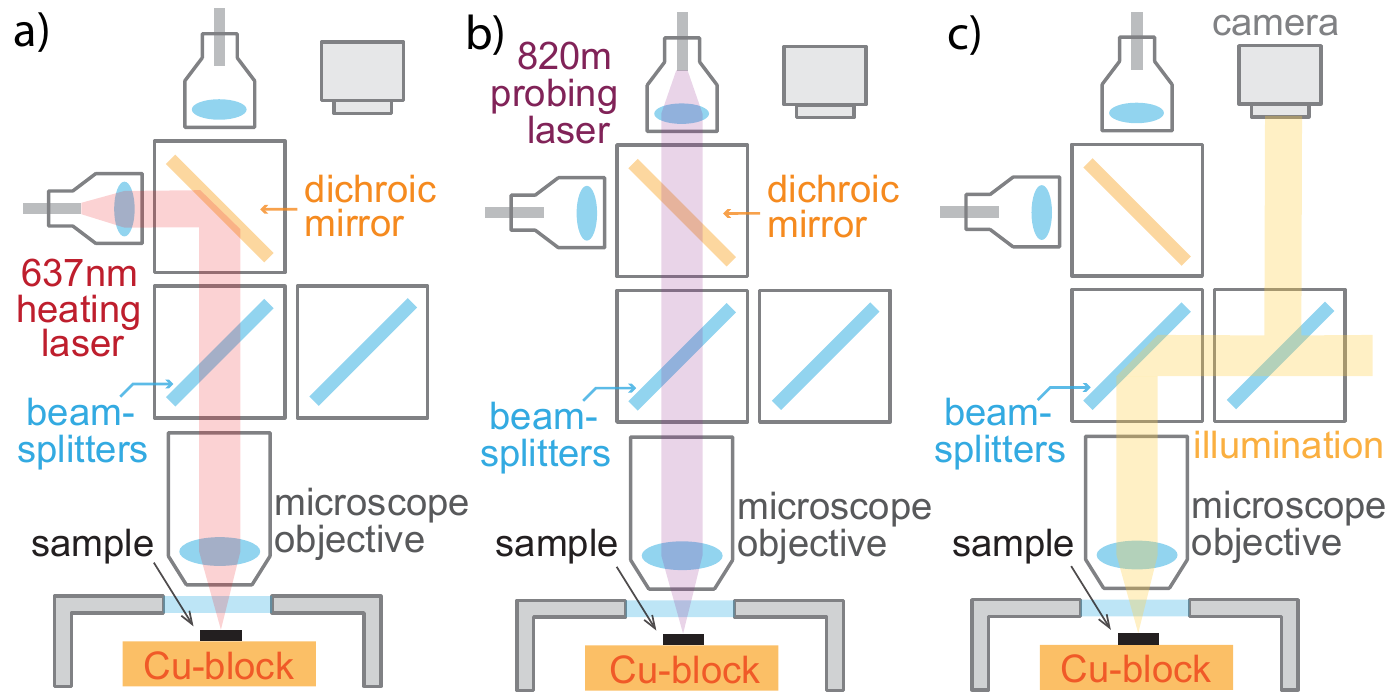}
	\caption{\label{setup}(color online) The schematic shows the optical paths of the setup. (a) Path of the heating laser (b) Path of the probing laser. The reflected light traverse the same path before gathered by a photodetector. (c) Path of camera vision. Cross-polarized picture is obtained by polarizing the incoming illumination and placing an analyzer in front of the camera.}
\end{figure}

\begin{figure}[ht]
	\centering
	\includegraphics[width=\linewidth]{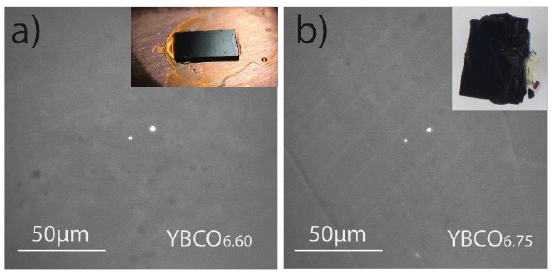}
	\caption{\label{sample} (color online)  Cross-polarized image of the samples showing both laser spots: left (smaller spot) is the 637nm heating spot and right (larger spot) is the 830nm probe laser spot.  a) a typical area of single crystal detwinned YBCO$_{6.60}$,  and b) a typical area of single crystal detwinned YBCO$_{6.75}$. Thin remnant strips of the opposite structural domain is visible as white lines, but have been shown to have negligible effect on the anisotropy measurements.  Insets show pictures of the crystals measured.}
\end{figure}

For the high resolution thermal diffusivity measurements we use a home-built photothermal microscope \cite{FantonThesis,XDWu1993}.  The microscope views the sample through a sapphire optical window in a cryostat, with the sample mounted to a cold finger just under the window.  A schematic is shown in Fig.~S\ref{setup}. A heating laser at 637 nm and a probing laser at 820 nm are focused onto the sample surface by the microscope objective.  The focused spots have Gaussian radii of approximately 1$\mu$m and 2$\mu$m, respectively, due to the diffraction limit of the different wavelengths, and can be moved independently over the sample surface.  A camera allows us to align the spots and observe the sample surface nearby.  
Fig.S~\ref{sample} shows a camera view of the two beams on different surfaces that we examined.

The output power of the heating laser is modulated with a sinusoidal profile $P(t)=P_0 [1 + \sin(\omega_0 t)]$. The modulation frequency $\omega_0/2\pi$ has a typical range of 1 kHz - 50 kHz, much slower than the microscopic equilibration time which is on the order of picoseconds \cite{Li2015}. This means that the parameters extracted are all within the in the DC limit of linear response, and are independent of the modulation frequency itself. The probing laser is aimed at a spot a small distance (typically $10-20$ $\mu$m) away from the heating laser. The reflected light from the probing laser is diverted by an optical circulator and fed into a photodetector. The AC component of the photodetector signal is then fed to a lock-in amplifier referenced to the laser modulation and the amplitude and phase are measured. The large differential reflectivity $dR/dT$ of the YBCO samples (about $10^{-5}-10^{-6}$ K$^{-1}$) at the wavelength of the probe light allows us to study their thermal properties with minimal disturbance.

\subsection{Measuring Thermal Diffusivity}
The diffusive transport of heat is governed by the diffusion equation
\begin{equation}
\label{diffusioneq}
	\frac{\partial \delta T(t,\vec{r})}{\partial t}
-D\nabla^2 \delta T(t,\vec{r})
=\frac{q(t,\vec{r})}{c}
\end{equation}
where $\delta_T$ is the temperature disturbance above the ambient temperature $T$, $\vec{r}=\{x_1, x_2, x_3\}$ is the spherical radial coordinate given in terms of the euclidean principal axes $x_i$, $q$ is the absorbed power density, $c$ is the volumetric specific heat capacity, $D \equiv \kappa / c$ is the thermal diffusivity, and $\kappa$ is the thermal conductivity.  Note that $c$ and $D$ are themselves functions of $T$, but in the limit of weak heating $\delta T << T$, we make the approximation $c(T+\delta T)\approx c(T)$ and $D(T+\delta T)\approx D(T)$.  As mentioned in the main text, the temperature disturbance from both lasers is $\lesssim$1 K through out the temperature range $25\sim 300$ K, so this approximation is valid.

The modulated power of the heating laser causes ripples in the temperature profile at the sample surface, which may be measured by the probing laser.  It is useful to write the response in frequency space $\tilde{\delta T}(\omega,\vec{r})$, where
\begin{equation}
\delta T(t,\vec{r}\,)=\int\tilde{\delta T}(\omega,\vec{r})\exp(-i\omega t)d\omega 
\end{equation}
We model the focused heat source as a point source, $q(t,\vec{r}\,)= P_0 e^{-i\omega t}\delta^3(\vec{r})$.  This approximation is valid as long as the distance from the heating spot is much larger than the spot radius.  In a semi-infinite isotropic system, the temperature profile is spherically symmetric and takes the form
\begin{equation}
	\tilde{\delta T}(\omega,r)=\underbrace{\frac{P_0}{\kappa}
	\frac{1}{r}\exp \bigg(-\sqrt{\frac{\omega}{2D}}r\bigg)}_{\text{amplitude}}
		\underbrace{\exp\bigg(-i\sqrt{\frac{\omega}{2D}}r\bigg)}_{\text{phase}}
			\label{diffsol}
\end{equation}
Our measurement gives us the response at the modulation frequency $\omega_0$.

\begin{figure}[ht]
	\centering
	\includegraphics[width=0.8\linewidth]{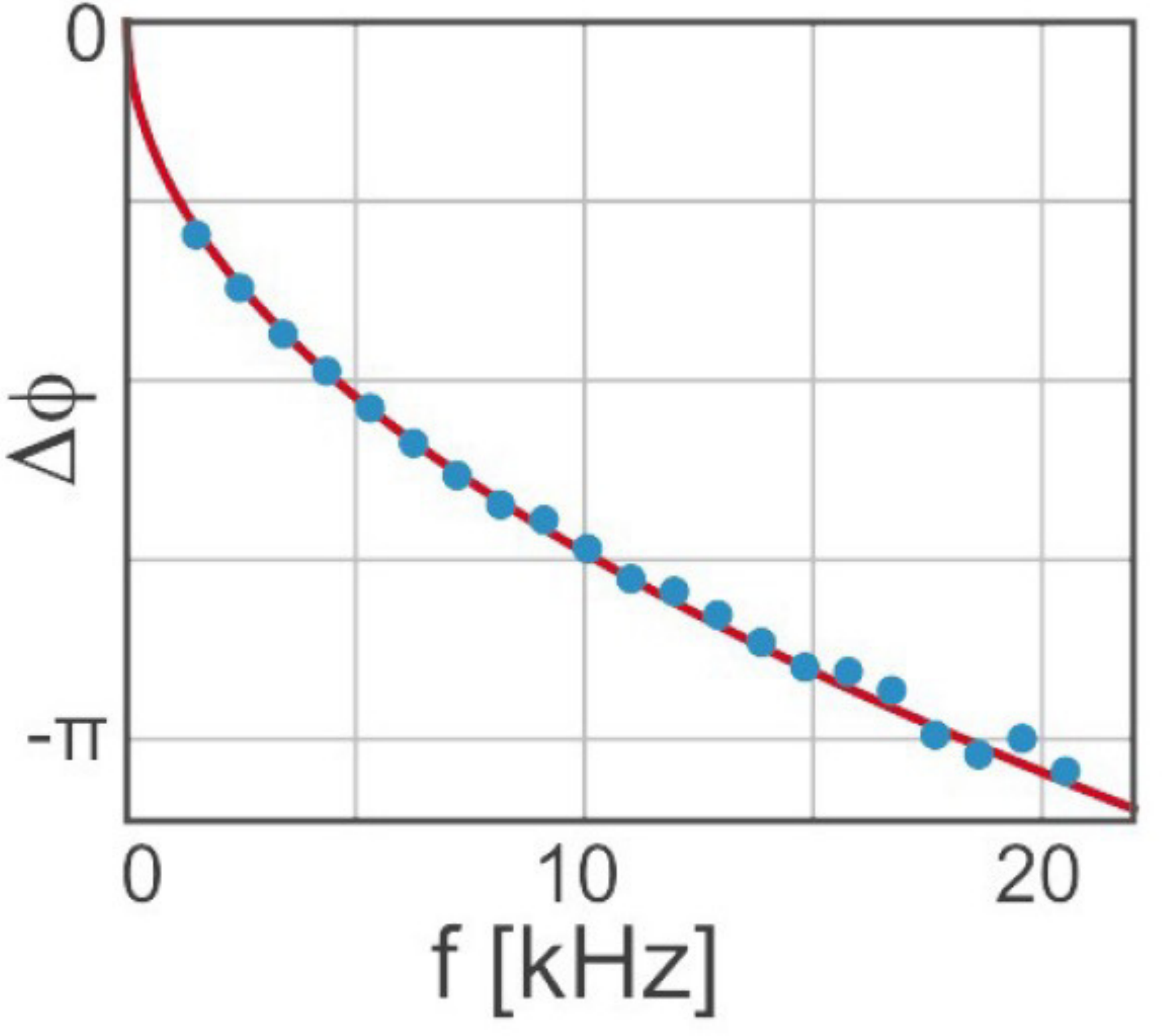}
	\caption{\label{phase}(color online) Typical phase delay data (blue circles) obtained by sweeping modulation frequency.  The red line is a fit using the form in Eq.~\ref{diffsol}.   }
\end{figure}

Although both the amplitude and the phase of the solution carry information about $D$, in actual measurements factors such as mechanical vibrations, fluctuations in the laser power, and the temperature dependence of the differential reflectivity $dR/dT$ affect the amplitude of the reflectivity oscillation.  These factors do not affect the phase of the signal, which is therefore the more robust probe. We obtain $D$ by fitting the phase delay $\phi$ between the source and the response signals as a function of $\omega$ at fixed $r$: $D = \omega r^2 / 2 \phi^2$. A typical fit is shown in Fig.S~\ref{phase}. 
It is important to note that  frequency sweeps at different distances yield consistent diffusivity values verifying the assumptions we used in obtaining the ideal solution.

\subsection*{Anisotropic Diffusion Equation}
A general system may be anisotropic, but can be reduced to the isotropic case through a change of variables.  To begin, we write down the heat diffusion equation for a generic system,
\begin{equation}
	\label{diffeqD}
	\frac{\partial \delta T(t,\vec{r})}{\partial t}
	-D_{ij}\nabla_i \nabla_j \delta T(t,\vec{r})
	=\frac{q(t,\vec{r})}{c}
\end{equation}
It is interesting to note that the dimension of the diffusivity consists only of space and time: $[D]=$ cm$^2/$s. A coordinate-isotropic solution can always be found through spatial rotation and scaling. Starting from the diagonal coordinates of the diffusivity tensor,
\begin{equation}
	\label{anio-iso}
	\sum_{ij} D_{ij}\nabla_i \nabla_j
	= \sum_i D_{i}\frac{\partial^2}{\partial x_i^2} 
	= D_1 \sum_i \frac{\partial^2}{\partial z_i^2}
	= D_1 \tilde{\nabla}^2
\end{equation}
 where we choose to keep the first axis untouched, and scale the rest accordingly, $z_{i}=x_i\sqrt{D_1/D_i}$. In the new coordinates, diffusion is an isotropic process, and diffusivity is characterized by a single number. In this new set of coordinate $\vec{z}$, the absorbed power density transforms accordingly as,
\[ q(t,\vec{z})=P_0 \sqrt{\frac{D_1^2}{D_2D_3 }} e^{-i\omega t}\delta^3(\vec{z}).\]
We now want to recover the anisotropic dependence on $D_{ij}$ on the surface of the sample in the original coordinate. Let $x_3=z_3=0$ on the surface, let $(\rho,\theta)$ be the original polar coordinate on the surface
\begin{equation}
	r^2=z_1^2+z_2^2=x_1^2+\frac{D_1}{D_2}x_2^2
	=\rho^2\big(\cos^2\theta+\frac{D_1}{D_2}\sin^2\theta\big)
\end{equation}
This leads to the exponent in Eqn.~\ref{diffsol} being rescaled as
\begin{equation}
	\sqrt{\frac{\omega}{2D}}r=\sqrt{\frac{\omega}{2}
	\frac{D_2\cos^2\theta+D_1\sin^2\theta}{D_1 D_2}}\rho
\end{equation}
which leads to the extracted diffusivity
\begin{equation}
D=\frac{D_1D_2}{D_2\cos^2\theta+D_1\sin^2\theta}
\label{anisotropy}
\end{equation}

\subsection*{Alignments and Data Collection}
To check the alignment confirm the orientation of the sample, we first measure diffusivity as a function of frequency for different orientations around the heating spot. This is typically done at room temperature and frequency sweeps are typically between 500 Hz and 20 kHz. For each orientation we generate a set of data similar to that shown in Fig.~S\ref{phase}, and from the fit extract the diffusivity $D(\theta)$. We then performed an overall fit of Eq.~\ref{anisotropy} to all sets of frequency sweeps at various sample orientations, where an additional offset angle is left as an additional free parameter. This offset is the difference between the arbitrary initial angle that we used and the actual principal directions of the crystal. All offset angles are later verified to align with the crystal edge within statistical uncertainties. Typical fits are shown in the insets of Fig.~1 of the main manuscript. We note that error bars are almost entirely due to an uncertainty of $\sim0.5~\mu$m in determining laser spots separation due to their finite spread. Excellent fit to the functional form Eq.~\ref{anio-iso} is obtained on both samples (see main text), including in the region on YBCO$_{6.75}$ crystal with thin stripes of remnant domains, Fig.~S\ref{sample}(b), verifying the anisotropic diffusivity measurements are effectively single-domain in nature.

The temperature dependent diffusivity on each axis is then taken by continuously very the temperature at a fixed laser separation on each axis, at a fixed modulation frequency ($\sim$20kHz) , and at the same region where the room temperature anisotropy has been measured.

\subsection*{Fits to the Electron-Phonon ``Soup'' Model}
We fit the thermal diffusivity measurement of the three YBCO$_{6+x}$ samples using the form
\begin{equation}
D = \frac{\hbar}{2k_BT} \left(\alpha_e \frac{c_e(T)}{c(T)}v_F^2 + \alpha_{ph} \frac{c_{ph}(T)}{c(T)}v_s(T)^2\right),
\label{soup}
\end{equation}
where literature values are used for the total heat capacity $c$\cite{YBCOcapacity}, the electronic heat capacity $c_e$\cite{Moler1997}, Fermi velocity $v_F$\cite{Fournier2010}, and sound velocities\cite{Lei1993,soundvelocity}.  The following table lists the coefficients $\alpha_e$ and $\alpha_{ph}$ in the plane ($a$) and chain ($b$) directions as extracted from the fits, with their standard errors:

\begin{table}[ht]
\centering
\caption{Results for fitting constants in equation~\ref{soup}}
  \begin{tabular}{  r | c  c  }
    				& $\alpha_e$		& $\alpha_{ph}$		\\ \hline
YBCO$_{6.60}$ plane 	& 0.14$\pm$0.01		& 2.0 $\pm$ 0.1 \\
    		 chain	& 0.413 $\pm$ 0.007	& 2.58 $\pm$ 0.06 	\\ \hline
YBCO$_{6.67}$ plane 	& 0.13 $\pm$ 0.02 	& 2.1 $\pm$ 0.3	\\
    		 chain	& 0.34 $\pm$ 0.06 	& 2.6 $\pm$ 0.5 	\\ \hline
YBCO$_{6.75}$ plane & 0.25 $\pm$ 0.02 	& 1.64 $\pm$ 0.09	\\
    		 chain	& 0.53 $\pm$ 0.02 	& 2.82 $\pm$ 0.2 	\\
  \end{tabular}
\end{table}

\subsection*{Preliminary Measurements on LSCO}
\begin{figure}[ht]
	\centering
	\includegraphics[width=\linewidth]{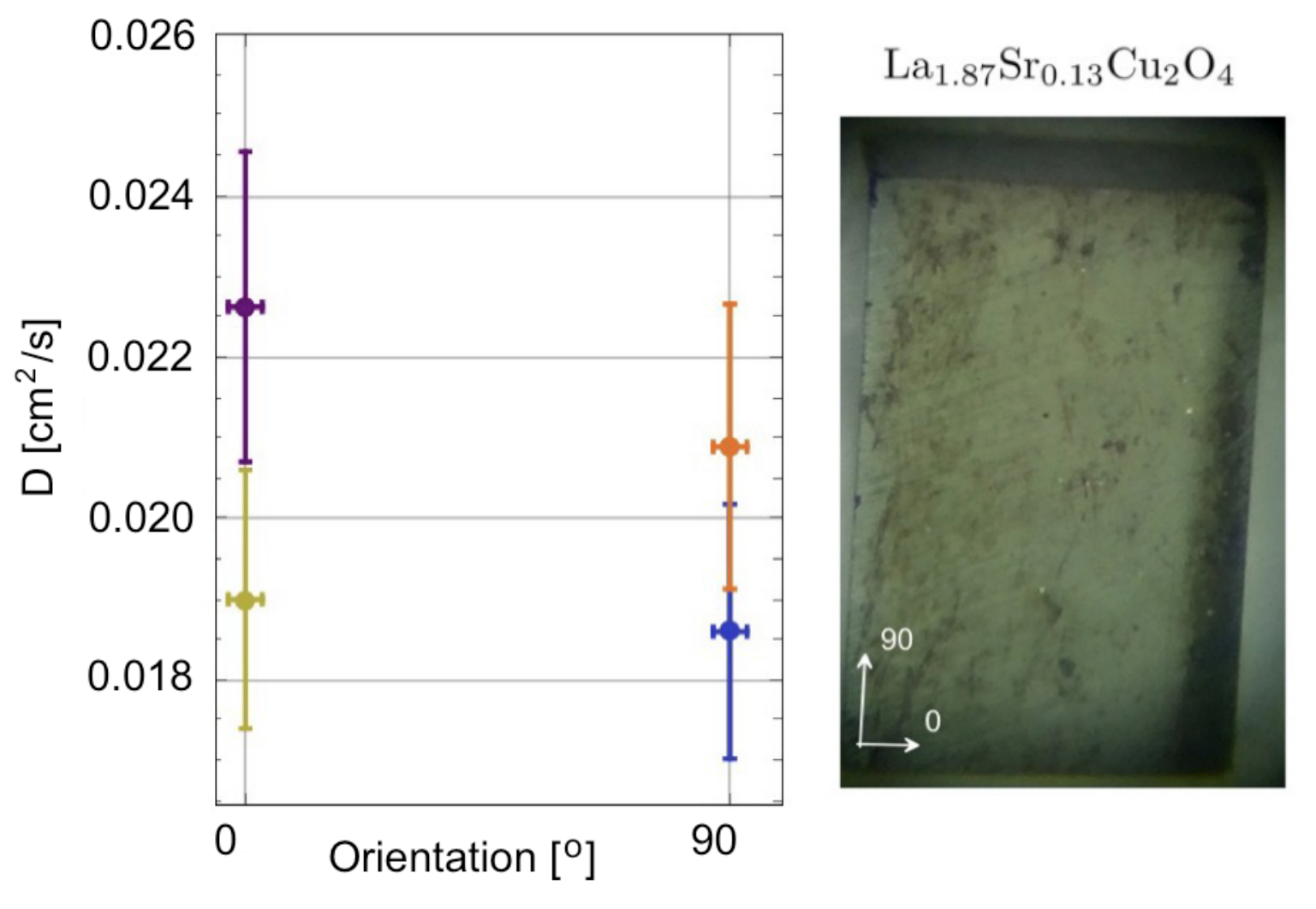}
	\caption{\label{lsco} (color online)  Left: Typical data in two perpendicular orientations measured on a La$_{1.87}$Sr$_{0.13}$Cu$_2$O$_4$ single crystal. Right: a picture of the surface of the crystal measured.  }
\end{figure}

To complete the diffusivity data that we present in Table 1 of the main text, we measured the thermal diffusivity on a $\sim1\times2\times 0.8$ mm$^3$ La$_{2-x}$Sr$_x$Cu$_2$O$_4$ ($x=0.13$) single crystal. In Fig.~S\ref{lsco} we show preliminary results of room-temperature measurements, together with a picture of the crystal. While the surface of the crystal was scratched, high-quality data could be obtained in high reflectivity regions. As expected, no obvious anisotropy was detected. 

\end{document}